# End-to-End Privacy for Open Big Data Markets

*Charith Perera (Open University), Rajiv Ranjan (CSIRO Digital Productivity Flagship), Lizhe Wang (Chinese Academy of Sciences)*

Abstract

The idea of an *open data market* envisions the creation of a data trading model to facilitate exchange of data between different parties in the Internet of Things (IoT) domain. The data collected by IoT products and solutions are expected to be traded in these markets. Data owners will collect data using IoT products and solutions. Data consumers who are interested will negotiate with the data owners to get access to such data. Data captured by IoT products will allow data consumers to further understand the preferences and behaviours of data owners and to generate additional business value using different techniques ranging from waste reduction to personalized service offerings. In open data markets, data consumers will be able to give back part of the additional value generated to the data owners. However, privacy becomes a significant issue when data that can be used to derive extremely personal information is being traded. This paper discusses why privacy matters in the IoT domain in general and especially in open data markets and surveys existing privacy-preserving strategies and design techniques that can be used to facilitate end to end privacy for open data markets. We also highlight some of the major research challenges that need to be address in order to make the vision of open data markets a reality through ensuring the privacy of stakeholders.

Introduction

Internet of Things (IoT) [1] promises to create a world where all the everyday objects (also called *things*) around us are connected to the Internet[1] and communicate with each other with minimum human intervention. The ultimate goal is to create '*a better world for human beings*', where objects around us know what we like, what we want, and what we need and act accordingly without explicit instructions. The Internet of Things allows people and *things* to be connected anytime, anyplace, with anything and anyone, ideally using any path/network and any service [2].

When examining the current IoT market place [3], it is clearly visible that we can broadly categorises the products and solutions into two segments. The majority of the products are aimed at individual customers (e.g. smart home owners) who may expect comfort and convenience through some kind of automation. For example, *WeMo* [3] is a Wi-Fi enabled switch that can be used to turn electronic devices on or off from anywhere. Another example would be *Nest* [3]. *Nest* is a thermostat that learns what temperatures users like and builds a context-aware personalised schedule to automatically control the household temperature efficiently. The other product group focuses on supporting business activities through collecting and analysing sensor data in enterprise and industrial domains. The potential clients for these products are mostly companies, not individual customers. For example, *Senseaware* [3] is a solution developed to support real-time shipment tracking. The context information such as location, temperature, light, relative humidity and biometric pressure is collected and processed in order to enhance the visibility of the supply chain. Another example would be *ParkSight* [3]. *ParkSight* is a parking management technology designed for cities. Context information is retrieved through sensors (magnetometers) embedded in parking slots.

---

[1] They may not be connected to the Internet directly but though intermediate devices.

Even though the distinction between these two categories can sometimes be vague, we can identify some unique characteristics. The main unique characteristic would be the target audience. In the first product category, potential clients are individual customers (i.e. families). As a result, ideally, the data generated by the products should belong to the individual product owners. In contrast, second product category is targeting enterprise customers. The data generated by this kind of solution may belong to the client company which bought the solution.

There are two important facts to highlight from the above discussion. Firstly, it is important to understand that different IoT solutions capture different types of sensor data in different contexts (e.g. households, factories, roads). Some IoT products may capture more private information (e.g. individual customer focused products) and others may capture less private information (e.g. enterprise or industry focused products). The second important fact is that, typically, these IoT products focus on achieving a single objective and data always move within the solution boundaries. Therefore, due to the fact that the data does not leave the product boundaries, the privacy risk related to these products are limited.

However, there is a significant amount of useful knowledge and insights that can be derived by combining, processing, and analysing the data collected by different IoT products [4]. It is more valuable if data collected by multiple data owners can be processed together. This kind of data sharing approach is are broadly referred to as sensing as a service [4]. Sensing as a service is the business model that drives the open data markets. However, despite the potential value of such data sharing and knowledge discovery, there is significant privacy risk involved in such approaches. This paper highlights the value of data sharing through open data markets powered by the sensing as a service model and while we provide design directions on how to ensure end to end privacy.

In rest of this paper, we briefly introduce the concept of sensing as a service and open data markets, followed by an analysis of privacy challenges associated. We discuss why sensing as a service model should be beneficial to everyone involved despite the privacy risks associated. Then, we survey and discuss some of the major privacy preserving design strategies towards addressing and mitigating those privacy risks, especially in IoT domain. Finally, we highlight some major research challenges that need to be addressed in order to build privacy protected open data markets.

<u>Vision towards Liberating Data</u>

This section provides a brief introduction to the sensing as a service model [4]. Sensing as a service is a business model which support data exchange between data owners and data consumers. Data owners purchase IoT products and deploy them in their own environments. These IoT products sense, analyse and perform actuation in order to make the data owners' lives easier. As a by-product, the collected data would be stored in an access restricted storage (usually referred as data silo). Data consumers are entities who would like to access other peoples' data for some reason. For example, a reason could be that a data analyst in an energy company may want to know how many energy inefficient legacy devices are used in a certain area. In this case, the data analyst is not interested in a particular household, but a whole set of households. We will discuss different use case scenarios later in this paper. When there are many data owners and potential data consumers, it creates an open data market. In this market data may not be freely available for anyone to access, only the meta-data would be. Meta-data would allow data consumers to understand what kind of data is stored in the silo by the data owners. Interested data consumers need to evaluate available meta-data data schemes and negotiate with the relevant data owners in order to get access to their data. The sensing as a service model utilizes the data primarily generated by IoT products.

Data collected by different IoT products has a significant value when aggregated and processed on a large scale (e.g. data collected from 10,000 households where each house has ten different IoT products). We discussed the details of sensing as a service model in [4], where different types of data owners, consumers, and mediator service providers are identified and analysed. Even though we have not yet discussed the privacy issues that could arise explicitly, you can imagine how privacy violation could occur in this kind of data sharing environment.

## Motivation for End to End Privacy Protection

In order to understand the significance of privacy challenges in the IoT domain, it is important to visualize how each concept presented so far would work in the real world. Figure 1 illustrates the use case.

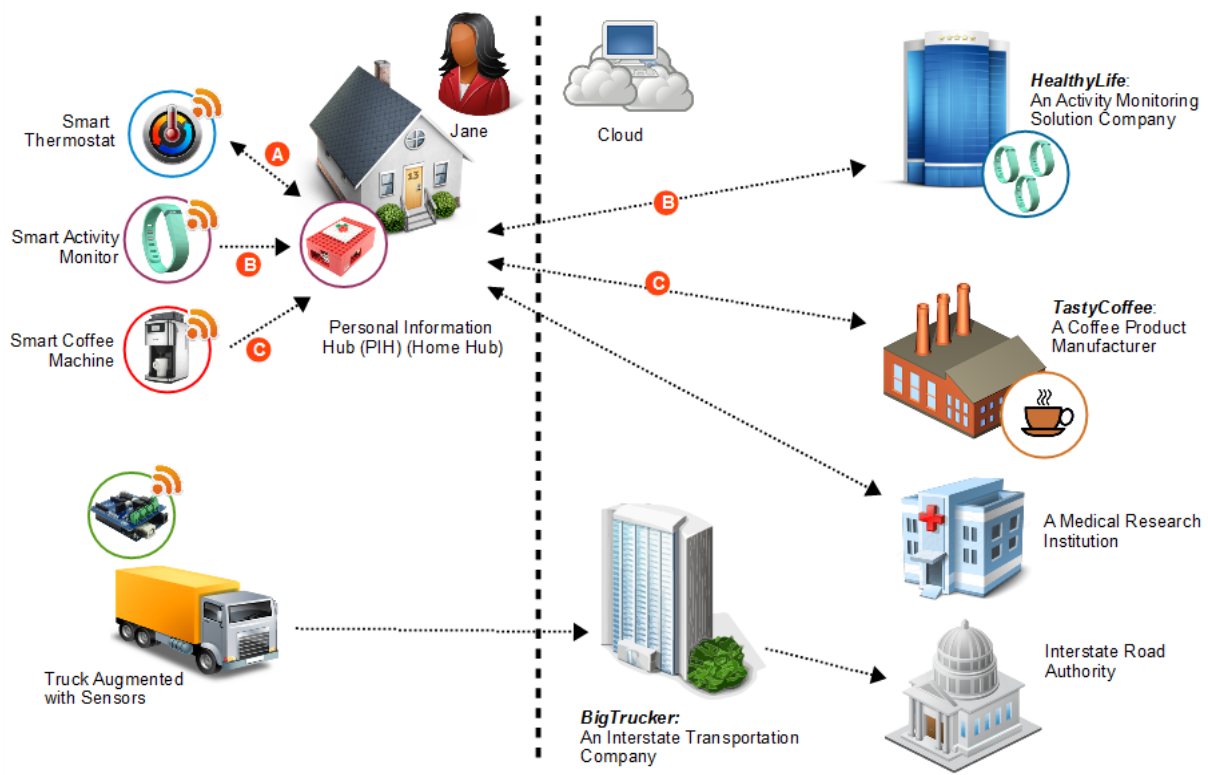

*Figure 1: Open data Market Supported by Sensing as a Service Model*

Let us introduce a persona we built to help with our discussion. *Jane* is a restaurant manager who works different shifts. She lives alone in her own house. She has purchased (and deployed) three different IoT products in her house. This first is a context-aware thermostat that controls indoor temperature based on user preferences. Secondly she has a smart coffee machine that automatically switches on and brews coffee when she gets up in the morning so by the time she arrives in the kitchen coffee is ready for her. Thirdly, *Jane* has bought a smart activity monitor that monitors her exercise patterns, food intake, step counts, goals, and so on. These three different products are purchased and deployed separately by *Jane* and they work independently.

There are different ways in which data could move within these IoT solutions based on their functionalities and user requirements. Let us consider IoT products such as smart thermostats. These products are trying to learn user preference over time and attempt to automatically actuate the heaters to control temperature. To do this kind of actuation, the data collected by the product does not need to leave the house itself. Therefore, the processing can be done by a small computer system built into the product (or using a Home Hub [3]). As a summary these products use its own sensors to sense the environment and process the data within the household. Then, they actuate the actuators to perform a certain tasks. We have denoted this kind of data flow in Figure 1 as (A).

Another type of data flow can be discussed using activity monitoring health kits. These category of IoT products use their sensors to sense the environment and do certain amount of processing and actuating (e.g. visualization and presentation, notification). However, for further processing, some part of data will need to be sent to the cloud services maintained by the product manufacturer. The reason and the advantage of such data flow is that IoT product manufacturers get to processes data retrieved from large number of users and give useful insights to the product owners in return. For example, if the data stays local, *Jane* will only be able to learn about her past, present and future result based on her own data which could be less useful. However, if *Jane* allows data to move to its manufacturer's service, she will be able to compare her performance in comparison to other similar users (e.g. same age, weight, height, job, workout patterns). Due to the fact that the IoT product manufacturers get to access data from a large number of users, they will be able to build more accurate and comprehensive prediction models to support not only *Jane* but also others as a whole community of users. As a result, here we can see the benefit that *Jane* would receive in return (e.g. monetary, coupon, points on a shopping card, etc.) for giving her data to the IoT product manufacturer. However, at the same time, we can see there are potential privacy risks involved in such data flows as well. We have denoted this kind of data flow in Figure 1 as (B).

In the sensing as a service model, we envision another type of data flow where data owners, like *Jane*, may give access to their data to a third party other than the respective IoT product manufacturer. We have denoted this kind of data flows in Figure 1 as (C). As presented in Figure 1, *TastyCoffee* is a manufacture of coffee products. They are keen to know how people like *Jane* consume coffee (e.g. patterns, amounts). *TastyCoffee* want to know whether there are any external factors that influence coffee consumption such as weather, temperature, workout patterns, etc. For example, *TastyCoffee* would like to discover any consumer patterns (e.g. *whether people tend to drink coffee before a workout*). Currently, the only way that they could discover this kind of information is through user surveys and focus group studies. However, such methods are time consuming, less accurate and expensive to carry out. However, if *TastyCoffee* can access *Jane's* silo (also thousands of other similar users) which consists of data recorded from all three of her IoT products (i.e. smart thermostat, smart coffee machine, activity monitoring products), they will be able to understand *Jane* (also thousands of other similar users) better and optimize their product supply chain. Such optimization will allow *TastyCoffee* to reduce their costs and wastage, which would increase their profits.

Further, such data will help *TastyCoffee* to improve their product lines and introduce new products to the market rapidly, which will also lead to strengthening of their brand value. Due to the additional value that *TastyCoffee* may generate, it can offer a return to the data owners to motivate them to give access to their data. From *Jane's* perspective, additional return would motivate her to trade her own data not only with *TastyCoffee* but also with other interested parties. However, this kind of data trading creates more privacy risks than the other two methods presented earlier.

In the scenario of *TastyCoffee*, the data will be traded based on commercial interests. However, data trading in the sensing as a service model could occur in a non-profit way as well. For example, a medical research facility may be interested in accessing the same data as *TastyCoffee*, but with the intention of conducting research into people's wellbeing by analysing correlation between coffee consumption, exercise patterns, weather, and indoor temperature. In this kind of scenario, the medical research centre would not be able to give a return in term of financial means, but they can use the research results to come up with actionable advice (e.g. consuming more than four cups of coffee reduces the impact of exercise by 20%[2]) and return them to the data owners as a return.

Let us consider another example that involves IoT products that we initially categorized as enterprise and industrial solutions. *BigTrucker* is a distribution company that handles goods on behalf of their clients (e.g. transport goods in between states). Their trucks are augmented with sensors and they sense the environment periodically and report back to the *BigTrucker* management centre. *BigTrucker* is using this IoT solution to monitor the health of employees (e.g. work condition over time) and status of the vehicles (e.g. maintenance estimation) and the quality of the goods transported. However, interstate road authorities may be interested in accessing this data to understand the environmental pollution and road conditions. Such data will help the authorities understand any environmental issues or infrastructure maintenance issues that need to be addressed urgently. Instead of deploying their own sensor networks and installing solar based power supplies, authorities may request data from *BigTrucker*. In return, *BigTrucker* may receive financial compensation. In this scenario, data is traded between two parties, however the privacy risks involved are lower due to the public and industrial nature of the data.

In the above mentioned scenarios, we explained why the sensing as a service model is important and how it can generate value for stakeholders. Further, we highlighted that the privacy risks associated with data trading vary significantly from one scenario to another based on the parties involved and the data traded. Personal data trading has more privacy risks in comparison to enterprise data. Further, when the amount of data being traded increases, the privacy risks are also increased. Similarly, more data allows the data consumers to derive more insights and generate more value out of it. Such potential value creation allows data consumers to provide a return to the data owners to motivate the trading of their data. Therefore, the responsibility of technology is to support data trading in open data markets while protecting the privacy of all stakeholders. This is the technology challenge we are facing today. In the rest of this paper, we survey existing privacy-preserving strategies and design techniques that can be used to facilitate end to end privacy for open data markets.

## Technologies for Privacy Preservation

So far we discussed why data trading between different parties is important and how such activities can create significant value to all the stakeholders involved. At the same time, we implicitly highlighted why the privacy risk involved in such data trading is high. In this section, we discuss how we can ensure that stakeholder privacy is protected when trading data by using existing privacy-preserving strategies and design techniques.

### Definition of Privacy
Before outlining survey privacy protection strategies and design techniques details, let us discuss *'what is privacy'* in brief. Privacy is a concept in disarray, which is difficult to articulate. "*Privacy is far too vague a concept to guide adjudication and lawmaking, as abstract incantations of the importance of 'privacy' do not fare well when pitted against more concretely stated countervailing interests*" [5]. One widely accepted definition, presented by Alan F. Westin [6], describes information privacy as *"the claim of individuals, groups or institutions to determine for themselves when, how, and to what extent*

---

[2] This is not medical advice based on any scientific results. This is an entirely made-up fact that we used to illustrate how an actionable advice may look.

*information about them is communicated to others"*. Roger Clarke [7] has mentioned that *"privacy is the interest that individuals have in sustaining a 'personal space', free from interference by other people and organisations"*.

Sometimes privacy is explained with the help of different dimensions. Privacy of the person, privacy of personal behaviour, privacy of personal communications, privacy of personal data [7] are the four main dimensions of privacy. In the Oxford Dictionary privacy is defined as *"a state in which one is not observed or disturbed by other people"[3]*. More importantly, privacy has been identified as a human right by the European convention[4] as well as by the Universal Declaration of Human Rights[5]. Further, the Charter of Fundamental Rights of the European Union defines the "*respect for private and family life*" in its Article 7 and adds a specific article on "protection of personal data" in Article 8. Additionally, Article 12 of the Universal Declaration of Human Rights protects an individual from "*arbitrary interference with his privacy, family, home or correspondence*," and "*attacks upon his honour and reputation*"[6]. This evidence strongly justifies the need to protect user privacy while we are attempting to harness the power of data trading and knowledge discovery to generate stakeholder value.

In parallel to the security protection goals, three goals have been proposed as privacy protection goals, namely *unlinkability, transparency,* and *intervenability* [8]. *Unlinkability* explains that data should not be combined from multiple data sources in such a way that together they would violate user privacy. *Transparency* means that stakeholders need to be informed about the data life cycle and what happens to each data item over time. This can be achieve through both technical and non-technical means such as auditing, laws, regulations, etc. The data owners should know what type data will be accessed, what kind of data sources will be combined, where the data will be processed, what kind of analytics will be used, what kind of results would be generated, and so on. A step going forward, *intervenability* says that data owners should be able to intervene at any time during the data life cycle so they can withdraw or change their consent over time. More importantly, data owners should have control over their data.

<u>Phases in Data Life Cycle</u>
During the life cycle, data moves through different phases. The phases are illustrated in Figure 2. It is important to note that these phases are somewhat vague in the real world and the order could be changed based on a given context. Today, IoT data processing is moving from cloud computing to fog computing. Fog Computing [9] is a paradigm that extends cloud computing and services to the edge of the network. Similar to cloud, fog provides data, computation, storage, and application services to end users. The distinguishing fog characteristics are its proximity to end users, its dense geographical distribution, and its support for mobility. There are advantage in processing data at the edge device. It avoids data communication and networking costs. Further, potentially, fog computing could reduce the potential privacy violation (e.g. processing smart home data within the house itself). However, the disadvantages are that edge devices may have limited computational capacity, limited energy, and more importantly limited data and knowledge about a given context. In order to derive more insightful and useful knowledge, data may need to be combined and processed together. Therefore, in IoT, data processing location is a balancing act.

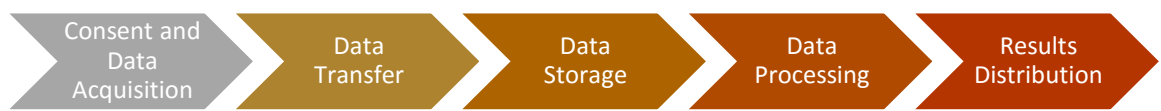

*Figure 2: Phases in Data Life Cycle*

---

[3] http://www.oxforddictionaries.com/definition/english/privacy
[4] http://conventions.coe.int/Treaty/Commun/QueVoulezVous.asp?NT=005&CM=7&DF=11/12/2014&CL=ENG
[5] http://www.un.org/en/documents/udhr/
[6] http://www.humanrights.com/what-are-human-rights/videos/right-to-privacy.html

In Figure 3, we illustrate a range of devices with different capabilities. We have grouped some commonly used devices in the IoT domain into a few different categories. Please note that this categorization is not done formally using any strict criteria. However, it approximates the differences between different groups in terms of the capabilities of the devices. The devices belonging to each category have different capabilities depending on processing, memory, and communication. They are also different in price where devices become more expensive towards the left of the figure. The computational capabilities also increase towards the left. Cloud computing is represented by Category 6 and rest of the categories may act as edge devices based on the context.

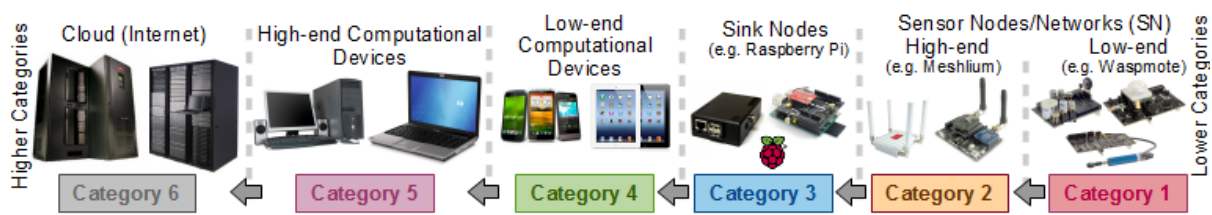

*Figure 3: Categorization of IoT devices based on their computational capabilities.*

As may now be apparent, sometimes data transfer, storage, and data processing could happen iteratively as the data moves from right to left. However, the technologies behind those phases would remain broadly the same. Therefore, in this paper we combine them into the above mentioned phases, despite the fact their actual execution sequence may vary depending on the formation of the fog network in a given context.

Privacy Preserving Strategies and Design Techniques
Hoepman [10] has proposed a number of privacy preserving strategies and design techniques. These techniques are mostly valid in IoT domain as well. Here we briefly introduce those strategies from an IoT perspective, referring to different situations.

The *Minimise* [11] design strategy says that the data consumers should only ask for the minimum amount of data that is required to achieve their objective. Typically, when the data consumers ask for more data, it creates more risk for the data owners. As a result, data owners may be reluctant to trade their data. Additionally, data owners may expect a higher return in order to match the additional risk involved. This design strategy comes into play in the *consent and data acquisition* phase. In the sensing as a service domain, negotiation will need to take place order to reduce the amount of data that is being traded between parties by considering associated risk and rewards. For example, if *TastyCoffee* wants to identify any pattern of coffee consumption and weather, they should not request any data related to motion sensors deployed in Jane's house. The smart coffee machine may communicate with motion sensors to identify whether Jane is awake. However, such information has no value to *TastyCoffee.* Further, anonymization (e.g. remove identity information) and use of pseudonyms (e.g. remove identity and introduce as resident of Milton Keynes) can also be used to minimise the amount of data traded [12].

A pseudonym is an identifier of a subject other than the subject's real name. Onion routing [13] is a technique for anonymous communication over a network. The sender remains anonymous because each intermediary node knows only the location of the immediately preceding and following nodes. This techniques can be used to perform anonymizing aggregation over large number of households. Instead of requesting data from large number of households and conducting the aggregation in a centralized location, onion techniques can be used to anonymously aggregate data on the fly.

Another design strategy is *informal*. It recommends embracing transparency and openness. This strategy is also relevant to the *consent and data acquisition* phase. However, information about other phases will be required to build a profile for both data owners and data consumers. Profiling is one of the most important tasks in open data markets as it help to conduct the data trading negotiations. Data owners should be informed about which data is processed, for what purpose, and by which means. It is important to let the data owners know about the ways the information is protected, and being transparent about the security of the system. This information will have a direct impact on data owner preferences to trade with a particular data consumer. As there would be risk and reward involved, trust plays a significant role in negotiating a particular trading between a given data owner and data consumer. Approaches similar to Privacy Preferences[7] (P3P) can be used to model the data owners' privacy preferences that may include their expectations about potential data consumers and their characteristics (e.g. level of trust, security, and openness of the techniques used in different phases of the life cycle).

*Hide* is a design strategy that recommends hiding data from plain view. This strategy is useful in both *data transfer* and *data storage* phases. Different types of encryption techniques [13] can be used during the data transmission from edge devices to cloud devices. Data may be stored in different types of device along that way as necessary. The encryptions that are supported by each device could vary depending on the computational capabilities of the device. Today, most of the time encryption techniques are employed in *data transfer and data storage* phases. However, recently homomorphic encryption techniques [14] have been introduced as a potential method to conduct computations over encrypted data. When homomorphic encryption is used, data is not required to decrypt in order to process. Homomorphic encryption techniques [14] can be incorporated with onion routing [13] to support end-to-end privacy and security. For example, individual data silos may generate results based on the data consumers' requests and the result would be passed from one silo to another where each silo may append its results to the incoming result using homomorphic encryptions. In this way, each silo may know about its own results but will have no knowledge about the incoming data.

The *Separate* strategy recommends storage of data in a distributed manner. In the IoT, this is the default assumption. Data owners may store their data in personal silos where they will give access to data consumers as part of the trading process. This strategy is mostly related to the *data storage* phase but also relevant to the *data processing* phase. There has been a substantial amount of research done on distributed data storage. Mostly this storage is called Personal Information Hub (PIH) (e.g. Hub of All Things[8], Lab of Things[9]). These edge devices specifically sit inside the data owner's home. Broadly, there are two methods by which PIHs may handle data processing. In one way, PIH does not allow data to move outside its physical boundaries (e.g. *Dataware* [15]). They accept a data analytical component into PIH and allow it to perform data processing tasks within the PIH boundaries. Only the result will be sent out from the PIH. In the other method, data is considered as movable and a limited amount of raw data will be sent out of the PIH. Data may then move either to other silos or to the centralized cloud over the fog network where data may be processed.

---

[7] http://www.w3.org/P3P/
[8] http://hubofallthings.com/
[9] http://www.lab-of-things.com/

*Aggregate* is another design strategy that is more related to the *data processing* phase. This strategy recommends the release of only the aggregated results from data silos. Typically, data becomes less sensitive if the data is sufficiently coarse grained, and the size of the group over which it is aggregated is sufficiently large. There could be a number of different ways to aggregate data. For example, data can be aggregated within the PIH. In our previous example, instead of returning raw data to the data consumers, PIH may return results saying the data owner has used the coffee machine five time per day on average over the last three months (i.e. aggregate over time). Such aggregated results do not provide detailed information about the coffee machine usage. Another aggregation method would be based on location. A potential result after distributed processing of multiple PIHs would be '*40% of Milton Keynes households use energy inefficient microwaves*'. Aggregation is a tricky task. For example, too much aggregation could hinder the knowledge discovery process and data consumers will not be able to derive useful knowledge. On the other hand, giving less aggregated data could be too risky for data owners where data consumers would be able to derive sensitive information about user behaviours and work patterns. Therefore, it is a challenging task to balance the ideal level of aggregation. Techniques widely used in this privacy-preserved aggregation are k-anonymity [16] and differential privacy [17].

*Control* is a design strategy which suggests that data owners should have the rights and access to the necessary tools to manage the data they trade to the data consumers. Again, this strategy is tricky due to the fact that some-times once data owners release results, it may not be possible to facilitate control functionalities that allows data owners to alter or remove their released data (i.e. results). Therefore, *Control* in IoT domain would be much limited compared to privacy protection in traditional banking or healthcare domain. Specifically, if the PIHs are releasing aggregated and processed data, facilitating control would be an impossible task. However, control strategy is significantly valid in early phases where the data owner gets to choose which data to trade to which data consumers under what circumstances, and so on. Further, even after the data trading negotiations are done and contracts are put it place, data owners should be able to change or cancel the contracts at any time.

The other two design strategies, namely, *Enforce* and *Demonstrate* are mostly non-technical in nature that would potentially cover all different phases of the data life cycle. *Enforce* recommends privacy policies to be compatible with legal requirements. *Demonstrate* recommends establishing a data controller to be able to demonstrate compliance with the privacy policy and any applicable legal requirements. This controller should be an independent third party organization which may examine a given technology system (e.g. a given data consumer) and evaluate, audit and log its behaviour and level of compliance towards privacy policies.

## Research Challenges and Future Direction

Though there are many research challenges in privacy preserving data analysis in the IoT domain, here we discuss three major challenges that need to be addressed towards realizing the vision of open data markets.

**Next Generation IoT Middleware for Data Analysis:** Since the 1990s there have been a number of guidelines proposed on designing and developing privacy preserving software systems. Privacy by Design [18] is a concept developed by Ann Cavoukian to address the ever-growing and systemic effects of information and communication technologies, and of large-scale networked data systems. Though these design principles are not specifically designed for the IoT domain, they encompass recommendations to build software systems that protect user privacy. Cavoukian proposed seven design principles 1) proactive not reactive (preventative not remedial), 2) privacy as the default setting, 3) privacy embedded into design, 4) full functionality (positive-sum), not zero-sum, 5) end-to-end security (full lifecycle protection), 6) visibility and transparency (keep it open), 7) respect for user privacy (keep it user-centric).

These design principles are still relevant in IoT domains as well. Further, the principles provide software designers, developers and architects some direction on how to realize the vision of open data markets. In addition to the people who are directly involved in building software, IoT envisions a strong community of data analysts who will be the force behind knowledge discovery. These are the people who are in charge of deriving knowledge and insights from large volumes of data. In the sensing as a service domain, they need to answer many questions on a daily basis such as *what kind of data to process*, *what kind of analytics need be used*, *where to get data from* and so on. While answering such questions, they also need to make sure that user privacy is respected at all times. This is a very challenging task, especially due to the variations in privacy preference of different data owners and their expectations. Further, accessing, transferring, storing, and processing data from each data source could require different privacy preserving technique to be employed. It would be impossible for the data analysts to handle such complexity manually. Therefore, we believe that there should be a middleware platform that allows data analysts to focus on data analysis and knowledge discovery tasks where the middleware autonomously (or at least semi autonomously) handles the usage of privacy-preserving techniques appropriately.

In the previous section, we discussed different techniques that can be used to preserve user privacy during different phases of the data life cycle. It may already be clear that there are multiple methods to perform a given knowledge discovery task based on a number of factors (e.g. moveability of data, computational capability of edge devices). The IoT middleware platform should be able to autonomously combine different privacy-preserving techniques in order to support end-to-end privacy. Additionally, the middleware platform will need to help the data analysts by providing useful tips (e.g. what kind of data is needed in order to discover certain knowledge or a particular pattern, what additional knowledge can be derived if more types of data are available, etc.) on which techniques to use if there is more than one possible way to accomplish a given task.

Conducting such composition tasks would be challenging to do manually especially due to the large number of possibilities. For example, developers may write new data analytics components that may allow discovery of new knowledge. The ideal IoT middleware should be able to analyse these new data analytical components and examine their potential impact towards user privacy and where such components can be run (e.g. on edge devices or in the cloud). Such IoT middleware will eliminate a significant burden on data analysts and it will also reduce the human error that could lead to user privacy violations.

***Consent Acquisition and Negotiation:*** In the IoT, user consent is about acquiring the required level of permission from users and non-users who are affected by the devices or services. In the traditional Web, the method of receiving user consent is through the privacy terms and policies presented to users through paragraphs of text. Recently, with the emergence of social media and mobile apps, consent acquiring mechanisms have changed. Researchers [19] have found that the current methods of asking user consent in social media platforms, such as Facebook are ineffective and most of the users underestimate the authorization given to third party applications. In some cases, developers may not provide accurate information to users for the consent decision. In other cases, developers may provide accurate information; however, the users would be unable to understand exactly what the consent entails through lack of technical knowledge.

In the sensing as a service domain one major type of user is data owners. Therefore, one of the major privacy challenges in the IoT, especially in relation to open data markets, is to develop technologies that request consent from data owners in an efficient and effective manner. This is a challenging task due to the fact that every data owner has very limited time and limited technical knowledge to engage in the process. The consent acquisition process is also part of the negotiation process. Research will need to combine principles and techniques of the human computer interaction and cognitive sciences. Further, the sensing as a service domain envisions that data consumers will request data from data owners. Sometimes it would be difficult of data owners to spend too much time on evaluating these data requests. Therefore, ideally, there should be a way to build privacy profiles of each data owner which encapsulate privacy preferences. Such profiling can be done by questioning data owners on their privacy preferences combined with user behaviour and data trading they perform over time. When a data request is received, autonomous systems will need to evaluate the request on behalf of the data owner in order to perform a preliminary filtering so it makes the data owner's life easier.

***Risk and Reward Modelling and Negotiation:*** After the preliminary filtering, the software systems on the PIH should provide the data owner with limited information that may include risk and reward analysis in relation to a given data trading task. Data owners should be able to understand the complete picture of what is going to happen to their data and what they will receive in return. Further, data owners should be able to negotiate with the data consumer regarding the amount of data to be traded and the related rewards. There are multiple ways to handle such negotiation which would could vary from manual negotiation (i.e. significant involvement of data owners) to autonomous negotiations. The data and consent acquisition should be a scalable process from both data owners' and data consumers' perspectives. Towards this, semi-autonomous and autonomous negotiation strategies will need to be developed which could consider factors such as data owners' preferences, how preferences have changed over time, data consumers' requirements, rewards, and so on. Modelling different privacy risks [20] and conducting negotiations is challenging task.

## Conclusions

Privacy protection is not only to be regarded as an individual value, but also as an essential element in the functioning of democratic societies. At the same time, open data markets that are expected to be created through the sensing as a service model have a significant potential to generate value to the society by reducing wastage, costs, and allowing more personalized services to customers. We first explained how sensing as a service could be beneficial to different stakeholders. We surveyed a number of privacy-preserving strategies and alternative design techniques that have been proposed in different domains and discussed them from the IoT perspective. During our survey, it was revealed that there are a number of research gaps in the field that need to be addressed in order to realize the vision of sensing as a service by creating open data markets. Future research efforts by the community will need to focus on addressing these research challenges.

Specifically, easy to use cloud based privacy-preserving data analytics platforms will enhance the ability of data analysts to focus on data analysis tasks instead of worrying about privacy violations. Developing novel techniques to advice, recommend and teach data owners about potential risks, threats, and rewards in the sensing as a service domain will encourage more data owners to participate in open data trading. From a non-technological point of view, incentive mechanisms in conjunction with strict auditing would help to preserve user privacy while supporting useful knowledge discovery.

**Author Biographies**

**Charith Perera** is a Research Associate at The Open University, UK. Currently, he is working on the Adaptive Security and Privacy (ASAP) research programme. He received his BSc (Hons) in Computer Science in 2009 from Staffordshire University, Stoke-on-Trent, United Kingdom and MBA in Business Administration in 2012 from University of Wales, Cardiff, United Kingdom and PhD in Computer Science at The Australian National University, Canberra, Australia. Previously, he worked at Information Engineering Laboratory, ICT Centre, CSIRO. His research interests are Internet of Things, Sensing as a Service, Privacy, Middleware Platforms, Sensing Infrastructure, Context-awareness, Semantic Technologies, Middleware, Mobile and Pervasive Computing. He is a member of both IEEE and ACM.

**Rajiv Ranjan** is a Senior Research and Julius Fellow at CSIRO, Canberra, where he is working on projects related to cloud and big data computing. He has been conducting leading research in the area of cloud and big data computing developing techniques for: (i) Quality of Service based management and processing of multimedia and big data analytics applications across multiple cloud data centers (e.g., CSIRO Cloud, Amazon and GoGrid); and (ii) automated decision support for migrating applications to data centers. He has published about 110 papers that include 55+ journal papers. He serves on the editorial board of IEEE Transactions on Computers, IEEE Transactions on Cloud Computing, IEEE Cloud Computing, and Future Generation Computer System Journals. According to Google Scholar Citations his papers have received about 3,300 citations and he has an h-index of 24.

**Lizhe Wang** is a Professor at Institute of Remote Sensing & Digital Earth, Chinese Academy of Sciences (CAS) and a ChuTian Chair Professor at School of Computer Science, China University of Geosciences (CUG). Professor Wang received his BE and ME from Tsinghua University and Doctor of Engineering from University Karlsruhe, Germany. He is a Fellow of IET, Fellow of the British Computer Society, and Senior Member of IEEE. Professor Wang leads a group at CAS on Spatial Data Processing and the High Performance Computing Lab at CUG.